\documentclass[11pt]{article}
\usepackage{hyperref}
\pdfoutput=1

\begin{document}
\title{How flowers catch raindrops}
\author{Guillermo J. Amador$^1$, Yasukuni Yamada$^1$, David L. Hu$^{1,2}$ \\
Schools of Mechanical Engineering$^1$ and Biology$^2$\\Georgia Institute of Technology, Atlanta, GA 30332, USA\\
}
\maketitle

Several species of plants have raindrop-sized flowers that catch raindrops opportunistically in order to spread their 0.3-mm seeds distances of over 1 m. In the following fluid dynamics video, we show examples of these plants and some of the high speed videography used to visualize the splash dynamics responsible for raindrop-driven seed dispersal. Experiments were conducted on shape mimics of the plants' fruit bodies, fabricated using a 3D printer. Particular attention was paid to optimizing flower geometries and drop impact parameters to propel seeds the farthest distance. We find off-center impacts are the most effective for dispersing seeds. Such impacts amplify the raindrop's speed, encapsulate seeds within drops, and direct the seed trajectory at angles optimal for long-distance dispersal.

\end{document}